\documentclass[12pt]{article}

\newcommand{\vtx}{\circle*{10}}

\begin{document}

\begin{center}{\Large{\bf
Multiple Deligne values: a data mine with\\[6pt]
empirically tamed denominators}}\\[6pt]
{\large David Broadhurst\footnote{
Department of Physical Sciences,
Open University, Milton Keynes MK7 6AA, UK;\\
Institut f\"{u}r Mathematik und Institut f\"{u}r Physik,
Humboldt-Universit\"{a}t zu Berlin}, \today}\end{center}

{\bf Abstract}:
Multiple Deligne values (MDVs) are iterated integrals on the
interval $x\in[0,1]$ of the differential forms $A=d\log(x)$,
$B=-d\log(1-x)$ and $D=-d\log(1-\lambda x)$, where $\lambda$
is a primitive sixth root of unity. MDVs of weight 11 enter
the renormalization of the standard model of particle
physics at 7 loops, via a counterterm for the self-coupling
of the Higgs boson. A recent evaluation by Erik Panzer
exhibited the alarming primes 50909 and 121577 in the {\em
denominators} of rational coefficients that reduce this
counterterm to a Lyndon basis suggested by ideas from Pierre
Deligne. Oliver Schnetz has studied this problem, using a
method from Francis Brown. This gave 2111, 14929, 24137,
50909 and 121577 as factors of the denominator of the
coefficient of $\pi^{11}/\sqrt{3}$. Here I construct a basis
such that no denominator prime greater than 3 appears in the
result. This is achieved by building a datamine
of 13,369,520 rational coefficients, with tame denominators, for the
the reductions of 118,097 MDVs with weights up to 11. Then numerical
data for merely 53 primitives enables very fast evaluation of all of
these MDVs to 20000 digits. In the course of this Aufbau, six
conjectures for MDVs are formulated and stringently
tested.

\section{Introduction}

The ever-burning motive\footnote{Dr Samuel Johnson (1709-1784)
remarked that {\em actions are visible, though motives
are secret}.\\Chapter~4 of~\cite{pam} offers an account,
by Kontsevich, but by not Zagier, of {\em motives} in mathematics.}
underlying this paper is the pursuit of a better
understanding of the number theory of the renormalization of
quantum field theory (QFT). Yet the reader need have no
acquaintance with QFT. It suffices to know that, in May 2014,
Erik Panzer achieved the remarkable feat of evaluating
the contribution of a notoriously difficult Feynman diagram~\cite{thesis}
to the beta-function for the quartic self-coupling of the
Higgs boson.

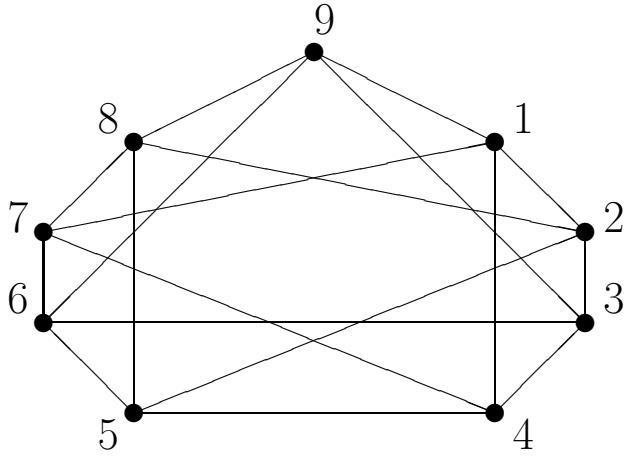
\begin{figure}
\begin{picture}(400,250)(-150,-30)
\put(0,0)\vtx\put(200,0)\vtx\put(250,50)\vtx\put(250,100)\vtx
\put(200,150)\vtx\put(0,150)\vtx\put(-50,100)\vtx\put(-50,50)\vtx
\put(0,0){\line(0,1){150}}
\put(0,0){\line(1,0){200}}
\put(200,0){\line(0,1){150}}
\put(-50,50){\line(0,1){50}}
\put(250,50){\line(0,1){50}}
\put(-50,50){\line(1,0){300}}
\put(0,0){\line(-1,1){50}}
\put(200,0){\line(1,1){50}}
\put(0,0){\line(5,2){250}}
\put(-50,100){\line(5,-2){250}}
\put(-50,100){\line(1,1){50}}
\put(200,150){\line(1,-1){50}}
\put(-50,100){\line(5,1){250}}
\put(0,150){\line(5,-1){250}}
\put(210,155){\makebox{{\Large 1}}}
\put(260,100){\makebox{{\Large 2}}}
\put(260,55){\makebox{{\Large 3}}}
\put(210,-20){\makebox{{\Large 4}}}
\put(-20,-20){\makebox{{\Large 5}}}
\put(-70,55){\makebox{{\Large 6}}}
\put(-70,100){\makebox{{\Large 7}}}
\put(-20,155){\makebox{{\Large 8}}}
\put(100,200)\vtx
\put(100,200){\line(1,-1){150}}
\put(100,200){\line(-1,-1){150}}
\put(100,200){\line(2,-1){100}}
\put(100,200){\line(-2,-1){100}}
\put(100,210){\makebox{{\Large 9}}}
\end{picture}
\caption{A symmetric graph with 9 indistinguishable
4-valent vertices on a Hamiltonian circuit and chords connecting vertices
whose labels are congruent modulo 3.}
\end{figure}
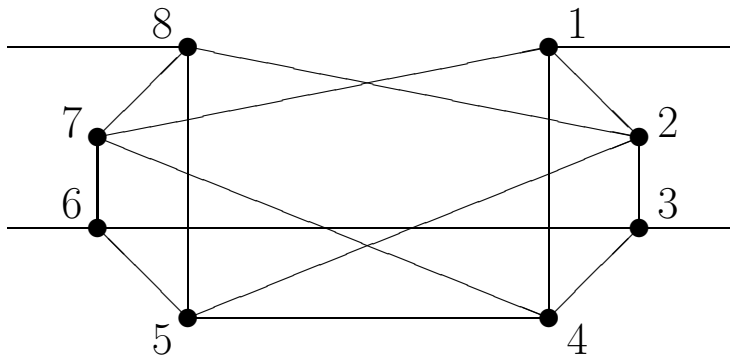
\begin{figure}
\begin{picture}(400,200)(-150,-30)
\put(0,0)\vtx\put(200,0)\vtx\put(250,50)\vtx\put(250,100)\vtx
\put(200,150)\vtx\put(0,150)\vtx\put(-50,100)\vtx\put(-50,50)\vtx
\put(0,0){\line(0,1){150}}
\put(0,0){\line(1,0){200}}
\put(200,0){\line(0,1){150}}
\put(-50,50){\line(0,1){50}}
\put(250,50){\line(0,1){50}}
\put(-50,50){\line(1,0){300}}
\put(0,0){\line(-1,1){50}}
\put(200,0){\line(1,1){50}}
\put(0,0){\line(5,2){250}}
\put(-50,100){\line(5,-2){250}}
\put(-50,100){\line(1,1){50}}
\put(200,150){\line(1,-1){50}}
\put(-50,100){\line(5,1){250}}
\put(0,150){\line(5,-1){250}}
\put(210,155){\makebox{{\Large 1}}}
\put(260,100){\makebox{{\Large 2}}}
\put(260,55){\makebox{{\Large 3}}}
\put(210,-20){\makebox{{\Large 4}}}
\put(-20,-20){\makebox{{\Large 5}}}
\put(-70,55){\makebox{{\Large 6}}}
\put(-70,100){\makebox{{\Large 7}}}
\put(-20,155){\makebox{{\Large 8}}}
\put(0,150){\line(-1,0){100}}
\put(-50,50){\line(-1,0){50}}
\put(200,150){\line(1,0){100}}
\put(250,50){\line(1,0){50}}
\end{picture}
\caption{Removal of any vertex from Figure~1 gives
a unique 7-loop subdivergence-free Feynman diagram whose
counterterm in $\phi^4$ theory is given by the MDVs in
$P_{7,11}$.}
\end{figure}

This beta-function determines the manner in which the
coupling changes with the energy-scale of the process. The
diagram in question may be obtained from Figure~1 by
removing one of the 4-valent vertices, leaving 4 half-edges
that correspond to external particles in the Feynman diagram of Figure~2.
The resultant
contribution to the beta-function does not depend on the
renormalization scheme and is given by a positive number
that does not depend on external momenta or internal masses.
In the census~\cite{census} of Oliver Schnetz this number is
referred to as the period $P_{7,11}$, since it comes from
the 11th in a list of 7-loop diagrams, by which is meant
that the first Betti number of these diagrams is 7.

So much for the important physics; now for the mathematics.

$P_{7,11}$ is a indeed a period~\cite{pam}, since it may be written as
a 13-dimensional integral of a rational function, over a
simplex with the rational boundary conditions that the
positive integration variables have a sum no greater
than unity, if one uses Feynman parameters.
Alternatively,
Schwinger parameters give the period as a projective integral. In
1995, using methods radically different from those of Feynman
or Schwinger, Dirk Kreimer and I obtained 11 good digits of
this period~\cite{pisa}. Since then, my numerical progress
has been modest: by 2011, I had obtained 21 good
digits of
\begin{equation}P_{7,11}=200.357566429275446967\ldots\end{equation}
For all other 7-loop diagrams in the census,
such precision is sufficient~\cite{BS} to discover an empirical
reduction to multiple zeta values (MZVs). Yet no credible
fit to MZVs was achieved in the case of $P_{7,11}$.

This was a problem crying out for analytical understanding.
Schnetz gave an argument~\cite{finite-fields},
based on counting the zeros of the
denominator of Schwinger's integrand in finite fields,
that $P_{7,11}$ might
evaluate to weight-11 polylogarithms of powers of the sixth root of
unity, $\lambda=(1+i\sqrt{3})/2$. In such a case, there
seemed to be little chance that 21 digits of numerical data
could be lifted to an exact analytical result. There are
$6^2 7^9=1,452,729,852$ legal 11-letter words in the
7-letter alphabet for iterated integrals of the differential forms
$d\log(x)$ and $-d\log(1- \lambda^n x)$, with $n=0,\ldots,5$.
The prospects seemed dim that a tiny subset of more than a
billion words could be chosen as suitable to fit $P_{7,11}$,
at merely 21-digit precision.

Panzer solved this problem, heroically, as part of his
larger work on a thesis~\cite{thesis} that implements and
extends, by computer algebra, Brown's
route map~\cite{route,route1} for reducing periods
to polylogarithms, when non-linearity of the denominators of
integrands may be avoided, in this case with considerable
cunning. Moreover, Panzer did this for many other Feynman
diagrams. As a notable Christmas present, he sent me in
December 2013 almost 1000 good digits of $P_{7,11}$. Yet
even this phenomenal precision was insufficient for either of
us to obtain an exact result in a basis that might,
with good luck, have dimension 144, the 12th Fibonacci number.

Undismayed, Panzer increased the precision of his
computations to 5000 digits and, by shrewd use of my
generalized parity conjecture~\cite{sixth}, obtained
an empirical reduction to a basis of size $144/2=72$,
thereby achieving a result which may be written, with
ingenuity, on a single page. But then a remarkable
phenomenon was observed. The rational coefficient of
$\pi^{11}/\sqrt{3}$ in his result for $P_{7,11}$ was
\begin{equation}C_{11}=-\frac{964259961464176555529722140887}
{2733669078108291387021448260000}\end{equation}
whose {\em denominator} contains 8 primes greater than 11,
namely 19, 31, 37, 43, 71, 73, 50909 and 121577. In the
normal course of events, this might be attributed to a poor
choice of basis. Yet Panzer had chosen what seemed, a
priori, to be a rather sensible basis, formed from Lyndon
words in much the same manner as I had done for alternating
sums~\cite{alt}, with consequent economy in a
datamine~\cite{mine} obtained with Johannes Bl\"umlein and
Jos Vermaseren.

To compound the puzzle further, Schnetz used Panzer's 5000
digits of data to obtain an alternative
formula~\cite{letter} that was even more bizarre. His
coefficient of $\pi^{11}/\sqrt{3}$ has a 48-digit denominator
that contains Panzer's 8 primes, above, and four new ones,
namely 47, 2111, 14929 and 24137. Yet Schnetz, like Panzer,
had been guided by what appeared to be sensible reasoning.

This paper is devoted to the twin tasks of understanding
the origin of such apparently gratuitous primes and of
building a datamine that avoids them. It proceeds as
follows.

Section~2 defines multiple Deligne values (MDVs), alerts the
reader to differing conventions for notating them and gives
a useful theorem. Section~3 gives 6 conjectures on MDVs.
Three of these are rather general and will be recognized as
quite intuitive by a reader familiar with MZVs and
alternating sums; the other three are much more specific and
concern novel features of MDVs that the author himself did
not suspect before undertaking empirical investigation.
Section~4 describes some of the large body of recent
evidence that supports these 6 conjectures. Section~5 is
entitled {\em the lure of Lyndon} and indicates why the
successful use of depth-length Lyndon words for alternating
sums is a dubious guide to follow in the case of MDVs. In
Section~6, I set about taming denominator primes, by an
Aufbau that is relatively straightforward at weights up to
9, becomes more demanding at weight 10 and plain difficult
at weight 11. Nonetheless, Section~7 gives a tolerably
compact result for the period $P_{7,11}$, in which no
coefficient has a denominator divisible by any prime greater
than 3. Section~8 explains the structure of a downloadable
datamine of 13,369,520 exact data for the 118,097 finite MDVs
with weights up to 11, together with high-precision
numerical data for merely 53 primitives, thus enabling very
fast evaluation of all these MDVs, to 20000-digit precision.
Finally, Section~9 offers comments and conclusions.

\section{Definitions, notations and theorem}

A {\em multiple Deligne value} is the evaluation, $Z(W)$,
as an iterated integral, of a word $W$ formed from letters
in the alphabet $\{A,B,D\}$ of the differential forms
$A=d\log(x)$, $B=-d\log(1-x)$ and $D=-d\log(1-\lambda x)$,
where $\lambda=(1+i\sqrt{3})/2$ is a primitive sixth
root of unity. Consider, for example, the word $W=DAB$. Then
\begin{equation}Z(DAB)\equiv
\int_0^1 \frac{\lambda\,dx_1}{1-\lambda x_1}
\int_0^{x_1} \frac{dx_2}{x_2}
\int_0^{x_2} \frac{dx_3}{1-x_3}\end{equation}
where it is important to note that I follow the ordering
of~\cite{ams,sixth,mine}, with the outermost integration
associated to the first letter of the word, in this case
$D$. This is sometimes called the {\em physics
convention}. Some mathematicians prefer to write their words
the other way round, with the outermost integration
corresponding to the last letter of the word.

If a word neither ends with $A$ nor begins with $B$, it is a {\em legal}
word, else it is a {\em bad} word, to which $Z$ can assign
no value, since the iterated integral diverges.
However it is often useful to manipulate bad words at intermediate
stages of a calculation, as will be shown.

The {\em weight} of $Z(W)$ is the length of the word
$W$, in this case $w=3$, and the {\em depth} of $Z(W)$
is the number of letters in $W$ that are not equal to $A$,
in this case $d=2$. The significance of depth become clearer
when one expands
\begin{equation}-d\log(1-\lambda^n x)
=\frac{dx}{x}\sum_{k>0}(\lambda^n x)^k\end{equation}
for $n=0,1$, in the case of MDVs, or more generally for
$n=0,\ldots,5$, in the case of words in the 7-letter alphabet
$\{A,B,C,D,\overline{D},E,\overline{E}\}$, where
$C=-d\log(1+x)$, $E=-d\log(1-\lambda^2x)$ and bars denote
complex conjugation, which gives $\overline{D} =-d\log(1-
\lambda^5x)$ and $\overline{E}=-d\log(1-\lambda^4x)$.

Every legal word of depth $d$, in the 7-letter alphabet,
has an evaluation as a $d$-fold nested sum of the form
\begin{equation}S{z_1,z_2,\ldots,z_d\choose
a_1,a_2,\ldots,a_d}\equiv
\sum_{k_1>k_2>\ldots>k_d>0}\prod_{j=1}^d \frac{z_j^{k_j}}{k_j^{a_j}}\end{equation}
where $z_j^6=1$ and $a_j$
is a positive integer. It is important to note that I
follow the convention of~\cite{bk,bgk,bbb,bbbl,ams,sixth,mine,PD1},
with the outermost
summation associated to the leftmost arguments, $z_1$ and
$a_1$, of the symbol $S$ for the nested sum. Thus, for
example,
\begin{equation}Z(DAB)=S{\lambda,\overline{\lambda}\choose1,2}
\equiv\sum_{m=1}^\infty\frac{\lambda^m}{m}
\sum_{n=1}^{m-1}\frac{\overline{\lambda}^n}{n^2}.\end{equation}
Again I remark that some mathematicians write things the
other way round, with the parameters of the outermost
summation at the end of their lists. As in the case of the
Lilliputian little-endians and the Blefuscan big-endians,
nothing is gained by arguing about which convention is
preferable. All that matters is that authors carefully
inform readers of the conventions adopted, so that results
are not mangled by mistranslation.

All iterated integrals are endowed with a {\em shuffle} algebra
\begin{equation}Z(U)Z(V)=\sum_{W\in{\cal S}(U,V)}Z(W)\end{equation}
where ${\cal S}(U,V)$ is the set of all words $W$ that
result from shuffling the words $U$ and $V$. Shuffles preserve
the order of letters in $U$ and the order of letters in $V$,
but are otherwise unconstrained. Thus, for example,
\begin{eqnarray}Z(AB)Z(CD)&=&Z(ABCD)+Z(ACBD)+Z(ACDB)\\&+&{}
Z(CABD)+Z(CADB)+Z(CDAB).\end{eqnarray}
The {\em full} 7-letter alphabet
$\{A,B,C,D,\overline{D},E,\overline{E}\}$ is also endowed with a
{\em stuffle} algebra, resulting from shuffling the
arguments of nested sums, taking account of the extra
stuff~\cite{ams} that results from terms when indices of
summation coincide. Thus, for example, the stuffle
\begin{eqnarray}Z(AB)Z(D)&=&S{1\choose 2}S{\lambda\choose1}=
S{1,\lambda\choose2,1}+S{\lambda,1\choose1,2}+S{\lambda\choose3}\\
&=&Z(ABD)+Z(DAD)+Z(AAD)\end{eqnarray}
may be combined with the shuffle
$Z(AB)Z(D)=Z(ABD)+Z(ADB)+Z(DAB)$ to prove that
$Z(DAD)+Z(AAD)=Z(ADB)+Z(DAB)$. One may also combine a stuffle
with a shuffle when one of the MDVs in the product is
divergent. For example, by equating the stuffle and shuffle
for the product $Z(AD)Z(B)$ one obtains the relation
$Z(ADD)+Z(AAD)=Z(ADB)+Z(ABD)$, from which the bad term
$Z(BAD)$ has been eliminated.

It is important to appreciate that the restricted alphabet
$\{A,B,D\}$ of MDVs is {\em not} endowed with a full stuffle algebra.
In most cases, the stuffle of a pair of MDVs does not give a
sum of terms each of which is a MDV, even though the
shuffle relation ensures that the total is expressible as a sum
of MDVs. For example, one may equate the stuffle and shuffle
for the product $Z(AD)Z(D)$ to obtain the relation
$Z(ADE)+Z(DAE)+Z(AAE)=2Z(ADD)+Z(DAD)$, which tells us that
the left-hand size is an {\em honorary} MDV, since it
is expressible in terms of MDVs. But that stuffle has told
us nothing new about relations between MDVs. This
is in marked contrast with the alphabet $\{A,B,C\}$
of alternating sums, which is closed under both
shuffles {\em and} stuffles.

To compensate for its lack of closure under stuffles,
the alphabet $\{A,B,D\}$ of MDVs has a beautiful
feature: the complex conjugate of a MDV is a MDV,
as will be shown after a few notational preliminaries.

{\bf Notation}: The map $Z$ operates on a word $W$ to produce a
value $Z(W)$ that is, in general, a complex number. It is
notationally convenient to extend $Z$ linearly, so that it
may act on sums of words with, in general, complex
coefficients. Thus the action on the empty word is $Z(1)=1$
and for a sum of words $T=\sum_j c_j W_j$ one obtains
$Z(T)=\sum_j c_j Z(W_j)$. I define a conjugate map
$\overline{Z}$ such that $\overline{Z}(T)$ is the complex
conjugate of $Z(T)$. One may abbreviate $n$ successive
occurrences of the same letter in a word by raising that
letter to the $n$th power, so that, for example, $A^2B D^3$
stands for $AABDDD$. Finally, I define the {\em dual},
$\widetilde{W}$, of a word $W$, to be the word obtained by
writing $W$ backwards and then interchanging $A$ and $B$, so
that, for example, the dual of $W=A^2B D^3$ is
$\widetilde{W}=D^3 A B^2$.

{\bf Lemma 1} [complex conjugation of MDVs]:\\
For any legal word $W$ in the $\{A,B,D\}$ alphabet,
$\overline{Z}(W)=(-1)^{n_D}Z(\widetilde{W})$,
where $n_D$ is the number of occurrences of $D$ in $W$.

{\bf Proof}: Map $x\to1-x$ in the iterated integral and use the
easily verified identity\\ $d\log(1-\overline{\lambda}(1-x))=
-d\log(1-\lambda x)$.

{\bf Lemma 2} [powers of $D$]:\\
Let $P_n\equiv(\pi/3)^n/n!$. Then $Z(D^n)=i^n P_n$.

{\bf Proof}: $Z(D)=-\log(1-\lambda)=i\pi/3$. The shuffle algebra
gives $Z(D)Z(D^{n-1})=n Z(D^n)$. Hence $Z(D^n)=i^n P_n$,
by induction.

{\bf Theorem 1} [sum rule at greatest depth]:\\
Consider the sum of words
$G_w\equiv\sum_{w>n>0}D^n B D^{w-1-n}$.
Then $2\Re Z(G_w/i^w)=(w-1)P_w$.

{\bf Proof}: Equate the shuffle and stuffle for the product
$Z(B)Z(D^{w-1})$, to eliminate the divergent term $Z(B D^{w-1})$.
This gives $Z(G_w)-Z(\widetilde{G_w})=(w-1)Z(D^w)$.
From Lemma 1, $Z(\widetilde{G_w})=(-1)^{w-1}\overline{Z}(G_w)$.
From Lemma 2, $Z(D^w)=i^w P_w$. Divide by $i^w$ to prove
the stated result.

{\bf Example}: At $w=3$, Theorem 1 gives $\Im Z(DBD+DDB)=-\pi^3/162$.
This will be needed in Section~4.

{\bf Remarks}: Lemma 1 greatly simplifies the problem of
reducing MDVs of a given weight $w$ to a basis. In the case
of alternating sums~\cite{mine} the datamine basis at $w=11$
has dimension 144. For MDVs of weight 11, needed for the
Feynman period $P_{7,11}$, I shall construct one basis of
dimension 72 for the real parts and another of dimension 72
for the imaginary parts. By halving the size of a basis,
one greatly extends the utility of integer-relation searches
based on numerical data, using the {\tt LLL} or {\tt PSLQ} algorithms
that are conveniently provided as options in the {\tt
lindep} procedure of {\tt Pari-GP}~\cite{pari}.

\section{Conjectures and remarks}

{\bf Conjecture 1} [Fibonacci enumeration]:
Let $F_n$ be the
$n$-th Fibonacci number. Then every {\bf Q}-linear combination of MDVs of weight $w$ is
reducible to a {\bf Q}-linear combination of $F_{w+1}$
basis terms between which there is no
{\bf Q}-linear relation.

{\bf Remarks}: A similar enumeration by Fibonacci numbers was
conjectured in 1996 for alternating
sums~\cite{alt}, constructed from words in the alphabet
$\{A,B,C\}$, where $C=-d\log(1+x)$. Conjecture 1,
for the $\{A,B,D\}$ alphabet, was inferred from a helpful
letter~\cite{letterp} by Deligne to the author in 1997, as reported
in~\cite{sixth}.

{\bf Conjecture 2} [enumeration of primitives]:
The dimension
$N_{w,d}$ of the space of primitive MDVs of weight $w$ and
depth $d$ is generated by
\begin{equation}\prod_{w>1}\prod_{d>0}(1-x^w y^d)^{N_{w,d}} =1-
\frac{x^2y}{1-x}\label{myc}.\end{equation}

{\bf Remarks}: By {\em primitive}, I mean irreducible to words of lesser
depth or their products. The generating function should
include $x^2y$, to record that the Clausen value $\Im Z(AD)$
is primitive. Then I suppose that $x$ stands for $(2\pi i)$
and that all else follows from the combination $(1-x-x^2y)$,
whose reciprocal, at $y=1$, generates the Fibonacci numbers
of Conjecture 1. In Conjecture 2, I divide by $(1-x)$,
to avoid assigning a depth to $(2\pi i)$. The corresponding
generating function for the dimension $E_{w,d}$ of the space
of primitives of weight $w$ and depth $d$ in the $\{A,B,C\}$
alphabet was conjectured in~\cite{alt} to be $(1-x^2-x y)/(1-x^2)$,
where $x y$ stands for $Z(C)=-\log(2)$ and
$x^2$ for $(2\pi i)^2$. Closed forms for both dimensions may
be obtained by taking a M\"obius transformation~\cite{alt,mine}
\begin{equation}T(a,b)=
\frac{1}{a+b}\sum_{n|a,b}\mu(n)P(a/n,b/n)\label{stephen}\end{equation}
of the binomial coefficients $P(x,y)\equiv(x+y)!/(x!y!)$ in
Pascal's triangle. The sum is over all positive integers $n$
that divide both $a$ and $b$; the M\"obius function,
$\mu(n)$, vanishes if $n$ is divisible by the square of a
prime and otherwise is $(-1)^{\omega(n)}$, where $\omega(n)$
is the number of prime divisors of $n$. Then Conjecture 2
gives $N_{w,d}=T(w-d,d)$, when $w>2d$. The corresponding
conjecture for alternating sums~\cite{alt,mine} is that
$E_{w,d}=T((w-d)/2,d)$, when $w-d$ is even and positive. I
shall comment in Section~5 on the relationship of the
symmetric array $T(a,b)=T(b,a)$ to the enumeration of Lyndon
words.

{\bf Conjecture 3} [generalized parity]:
The primitives of
Conjecture 2 may be taken as real parts of MDVs for which
the parities of weight and depth coincide and as imaginary
parts of MDVs for which those parities differ.

{\bf Remarks}: Conjecture 3 is a special case of a wider conjecture
of this kind, first made in the context~\cite{sixth} of the
7-letter alphabet formed from $d\log(x)$ and $-d\log(1-
\lambda^n x)$, with $n=0,\ldots,5$. Here I invoke a
generalized parity conjecture only for $n<2$, where it is
far easier to test.

{\bf Conjecture 4} [sum rule at odd weight]:
At odd weight $w>1$,
there exists a {\em unique} {\bf Z}-linear reducible
combination
\begin{equation}X_w=\sum_{k=1}^{(w-1)/2}C_{w,k} \Im Z(A^{w-2k-1}D A^{2k-
1}B),\end{equation}
with $C_{w,1}>0$ and integer coefficients $C_{w,k}$ whose
greatest common divisor is unity. Moreover, all of the
coefficients are non-zero, $X_w$ is free of products of
primitives and hence $X_w/\pi^w$ reduces to a rational number.

{\bf Remarks}: Conjecture 4 lies at the heart of the present
paper. From Conjecture 2 it follows that there should be
$(w-3)/2$ depth-2 primitives, at odd weight $w>1$, and from
Conjecture 3 that one may take these as the imaginary parts
of suitable MDVs. Thus there should be at least one
combination of the $(w-1)/2$ imaginary parts in $X_w$ that
is reducible to $\pi^w$ and products of depth-1 primitives.
Conjecture 4 asserts more than this, namely that
there is precisely one such combination, that every
coefficient $C_{w,k}$ is non-zero, that products of depth-1 primitives
do not occur in the reduction and hence that the result is a
rational multiple of $\pi^{w}$.
The price of such simplicity is high, since the
integers $C_{w,k}$ grow rapidly with $w$ and hence pose a
significant obstacle to constructing a sensible set of
primitives. By eliminating one of the $(w-1)/2$ imaginary
parts in $X_w$ one may end up dividing by a large prime
factor of one of the integers $C_{w,k}$. Moreover, this
problem may become worse if one chooses depth-2 primitives of
odd weight from words in the restricted alphabet $\{A,D\}$,
as was done by Panzer~\cite{thesis}. In Section~5, I shall
show how this introduced the prime 50909 into denominators
at weights 9 and 11 and the prime 121577 into denominators
at weight 11.

{\bf Conjecture 5} [honorary MZV at even weight]:\\
At even weight
$w$, the depth-2 real part $\Re Z(A^{w-2}D^2)$ is reducible
to MZVs.

{\bf Remarks}: Conjecture 5 refers to a single depth-2 MDV. So one
might suppose, at first sight, that it ought to be fairly
easy to prove. In fact, there is a significant obstacle to a
proof for all weights: the price that $\Re Z(A^{w-2}D^2)$
pays for being an honorary MZV is that its reduction to MZVs
may entail an increase in depth. Thus begins a tussle with
the intricacy of the Broadhurst-Kreimer (BK) conjecture for
the depth-graded enumeration $D_{w,d}$ of primitive MZVs, via our
infamous~\cite{FB1} generating
function~\cite{bk}
\begin{equation}\prod_{w>2}\prod_{d>0}(1-x^w y^d)^{D_{w,d}} =1-
y\frac{x^3}{1-x^2} +y^2(1-y^2)
\frac{x^{12}}{(1-x^4)(1-x^6)}
\label{bkc}\end{equation}
whose final rational function of $x$ also generates the
numbers $M_w$ of modular forms of weight $w$ of the full
modular group. The next conjecture circumvents the BK
problem of increase of depth, by including an alternating
sum for each modular form.

{\bf Conjecture 6} [modular forms and alternating sums]: For even
weight $w$, there exists a {\em unique} {\bf Q}-linear combination
\begin{equation}Y_w=3^{w-4}\Re Z(A^{w-2}D^2)+
\sum_{k=1}^{M_w}Q_{w,k}Z(A^{w-2k-2}C A^{2k}B),\end{equation}
with rational coefficients $Q_{w,k}$,
such that $Y_w$ reduces to depth-2 MZVs.

{\bf Remarks}: Conjecture 6 is notable for associating a set of
$M_w$ uniquely defined rational numbers to a set of modular
forms of the same cardinality. To motivate it, I invoke a
phenomenon called {\em pushdown}, first observed
in~\cite{alt} and later studied in detail using the MZV
datamine~\cite{mine}. Pushdown refers to the fact that some
MZVs regarded as irreducible in Don Zagier's MZV alphabet
$\{A,B\}$ are reducible to primitives of lesser
depth in the alphabet $\{A,B,C\}$ of alternating sums. This
occurs at weight 12, where the first modular form appears
and a depth-4 MZV has a pushdown to the alternating sum
$Z(A^8C A^2B)=\sum_{m>n>0}(-1)^{m+n}/(m^9n^3)$. Then
$Q_{12,1}=2^8$ records this fact rather compactly, since
$\Re Z(3^8A^{10}D^2+2^8A^8C A^2B)$ is empirically reducible
to depth-2 MZVs. In Section~4, I give all such rational
numbers up to weight 36, where 3 modular forms occur.

\section{Evidence}

The MDV datamine of Section~8 stands as strong witness for
Conjectures 1 to 5, which are in perfect accord with the
empirical reductions obtained for the 118,097 finite MDVs
with weights up to 11. Each of these MDVs was evaluated to
4000-digit precession, using the method in Section~7 of~\cite{ams},
devised with Jonathan Borwein, David Bradley and Petr Lisonek.
The datamine of Section~8 now extends the precision to 20000 digits.

This datamine records 13,369,520 non-zero rational
coefficients obtained in empirical reductions to putative
basis terms whose enumeration accords with the Fibonacci
dimensions of Conjecture 1. In no case was it necessary to
supply {\tt lindep} with more than 2300 digits, to obtain
these results. Thus the probability of a spurious
reduction is comfortingly less than $1/10^{1700}$.
Moreover, {\tt lindep} could discover no credible rational
relation between the elements of the putative basis,
at 4000-digit precision.
I make the obvious, yet sobering, remark
that a proof of Conjecture~1 would require, inter alia, a
proof that $\zeta_3^2/\zeta_2^3$ is not a rational
number, which seems to lie beyond the present intellectual
capabilities of humankind.

By combining Conjectures 2 and 3 one arrives
at a divide-and-conquer formula $F_n=F_n^++F_n^-$ that
splits the $n$-th Fibonacci number into
$F_n^\pm=(F_n\pm\chi_3(n))/2$, where the
character $\chi_3(n)=\chi_3(n+3)$ is 0 if $n$ is divisible
by $3$ and $\chi(\pm1)=\pm1$. Then the conjectured
dimensions for real and imaginary parts at weight $w$
are $D_R(w)=F^+_{w+1}$ and $D_I(w)=F^-_{w+1}$, respectively.
It follows that the generating function for real parts is
\begin{equation}G(x)\equiv\frac{1}{1-(x+x^2)^2}
=\sum_{w\ge0}D_R(w)x^w\label{DR}\end{equation}
giving sequence {\tt A094686} of the OEIS~\cite{oeis}:\\
{\tt 1, 0, 1, 2, 2, 4, 7, 10, 17, 28, 44, 72 $\ldots$}\\
with an initial entry $D_R(0)=1$ recording the empty word.
For the imaginary parts, the generating function is
\begin{equation}H(x)\equiv(x+x^2)G(x)
=\sum_{w\ge0}D_I(w)x^w\label{DI}\end{equation}
giving sequence {\tt A093040} of the OEIS~\cite{oeis}:\\
{\tt 0, 1, 1, 1, 3, 4, 6, 11, 17, 27, 45, 72 $\ldots$}\\
with an initial entry $D_I(0)=0$ recording that
the empty word evaluates to a real number, Z(1)=1.
It is rather satisfying that both sequences are generated
so simply using only the quadratic $(x+x^2)$ that
reminds us that $\Im Z(D)$ is a rational
multiple of $\pi$ and that $\Im Z(AD)$ is not believed
to be a rational multiple of $\pi^2$.

Conjectures 1, 2 and 3 are in accord with motivic
reasoning~\cite{PD1,thesis}, which establishes that
$N_{w,d}\le T(w-d,d)$. It seems to
me that it would be a waste of time trying to
falsify these conjectures at higher weights, $w>11$.
By contrast, Conjectures 4, 5 and 6 seemed to merit close
attention at weights greater than 11.

Conjecture 4 concerns the depth-2 imaginary parts
\begin{equation}I_{a,b}\equiv\Im Z(A^{b-a-1}D A^{2a-1}B)=
\sum_{m>n>0}\frac{\sin(\pi(m-n)/3)}{m^{b-a}n^{2a}}\end{equation}
with $b>a>0$ and odd weight $a+b$. At each odd weight $w>1$,
it asserts, inter alia, that there exists a {\em unique} vector
of non-zero integers, $C_{w,k}$ for $k=1,\ldots,(w-1)/2$,
with unit content, such that
\begin{equation}X_w=\sum_{w>2k>0}C_{w,k}I_{k,w-k}=Q_w P_w\end{equation}
where $P_w\equiv(\pi/3)^w/w!$ and $Q_w$ is a
rational number. I have checked this at
4000-digit precision, up to weight $w=11$, using {\tt lindep},
which revealed, at each odd weight, the existence of precisely one combination
of imaginary parts that is reducible to products
of depth-1 primitives and powers of $\pi$.
To my considerable initial surprise, {\tt lindep}
gave 0 for the coefficients of all products, leaving only $P_w$.
This singular circumstance will be pursued in Section~5.

The datamine uses $I_{a,b}$, with $b>a>1$ and odd weight $a+b$,
and the depth-1 primitives
\begin{eqnarray}
R_{2k+1}&\equiv& Z(A^{2k}B)=\zeta_{2k+1}
=\sum_{n>0}\frac{1}{n^{2k+1}},\\
I_{2k}&\equiv&\Im Z(A^{2k-1}E)={\rm Cl}_{2k}(2\pi/3)
=\frac{\sqrt{3}}{2}\sum_{n>0}\frac{\chi_3(n)}{n^{2k}},\end{eqnarray}
with $E$ used for the Clausen values,
since $2^n\Im Z(A^n D)=(2^n+1)\Im Z(A^n E)$
and the use of $D$ may induce unwanted denominator primes
such as $43|(2^7+1)$ and $19|(2^9+1)$, at weights 8 and 10.
Similarly, $B$ is preferable to $D$ for the real primitives,
since $2^{n+1}3^n\Re Z(A^n D)=(2^n-1)(3^n-1)Z(A^n B)$ and
thus the choice of $D$ may induce the denominator primes
13, 17, 31, 41 and 61, at weights less than 12.

A basis for reductions of imaginary parts of
depth-2 MDVs of odd weight $w>1$ has, according to Conjectures 2
and 3, dimension $(w-2)$, comprising $(w-3)/2$ primitives,
$(w-3)/2$ products of primitives and, finally, $\pi^w$.
There are $N=3w-4$ finite imaginary parts, namely
those of the words $A^j D A^k D$, $A^j D A^k B$, $A^j B A^k D$,
for $ j\ge0$, $k\ge0$ and $j+k=w-2$, with the bad $w$-letter word
$BA^{w-2}D$ omitted at $j=0$, in the last case.
Shuffles of products $Z(A^j D)Z(A^k D)$
and $Z(A^j B)Z(A^k D)$ provide us with $(w-1)/2+(w-2)$ relations,
since we should avoid the bad shuffle $Z(B)Z(A^{w-2}D)$. As explained
in Section~2, stuffles of $Z(A^j D)Z(A^k D)$ are useless,
since they take us out of the $\{A,B,D\}$ alphabet. There
are $(w-1)$ useful stuffles of $Z(A^j B)Z(A^k D)$, for which
the case with $j=0$ is now allowed, by subtracting the corresponding
shuffle, to eliminate the bad word. The tally of relations
is thus $M=(5w-7)/2$. For a given odd $w$, it is then a simple exercise
in computer algebra to generate the $M\times N$ matrix for these
relations and compute its rank deficiency, which was proven to
be $(w-1)/2$ for all odd $w\le31$.

Since this rank deficiency exceeds the conjectured number
of primitives by unity, {\em one of our relations is missing.}
That is why the reduction $X_w$ of
Conjecture~4 is so significant: it is the kernel, at $d=2$,
of the ineluctable deficiency of depth-restricted algebra
for MDVs, which suffer from a stuffle algebra that does
not close. The remedy is clear: one should use the
good features of MDVs, celebrated in the lemmas and theorem
of Section~2.

{\bf Example}: Set $w=3$. Then Conjecture 4 requires that
$X_3=I_{1,2}=\Im Z(DAB)=Q_3P_3$, for some
rational $Q_3$. This is {\em impossible} to prove
using only shuffles and stuffles of products of depth-1 MDVs.
But it becomes trivial to prove when one adds the result
obtained in Section~2 from Theorem 1, at depth 3,
namely that $\Im Z(DBD+DDB)=-\pi^3/162$. Then Lemma 1
converts this to a result that we were lacking at depth 2,
namely $\Im Z(DAD+ADD)=\pi^3/162$, and very
simple algebra gives the required rational $Q_3=\frac72$.

Now imagine trying to find the rational number $Q_{11}$
by pure algebra. Since Theorem 1 was needed at $w=3$,
one may reasonably suppose that it is also needed at $w=11$.
But Lemma 1 only lowers the depth by unity, to $d=10$,
and we are seeking a rational number at $d=2$. So it looks
as if we may need to do hefty algebra on the 77,708 weight-11 MDVs
with depths $d\le10$, using the even larger number of relations
between them given by shuffles, stuffles and duality.
Such a Herculean task might be achievable, using
Jos Vermaseren's programme {\sc Form}.
More economically, {\tt lindep} returns the empirical result
\begin{equation}
Q_{11}=841838813449645=5\times11\times809\times43627\times433673
\end{equation}
in the twinkling of an eye. It took a good while to obtain and factorize
\begin{eqnarray}Q_{31}&=&\frac{5}{7}\times 432650667045719
\times 101610941211668471750779 \times p_{49} \times p_{81}\\
p_{49}&=&1052453969156963777695781293476878259787114222411
\end{eqnarray}
where $p_{81}$ is the following 81-digit prime:
\begin{verbatim}
5398660771478298532475166018701166835343\
25958155228637043335803543859216008062953
\end{verbatim}
found by {\tt GMP-ECM}~\cite{ecm}, using the parameters
{\tt B1=3000000, sigma=2086811470}.

I have tested Conjecture~4 up to $w=31$ and found it to be
flawless. The datamine of Section~8 provides a list of
integer coefficients, $C_{w,k}$, and the corresponding
rationals, $Q_w$. The largest prime in the denominators of
the rationals $Q_w$, for odd $w\le31$, is merely 13, yet the
largest prime in the numerators has 137 digits.

An iterative method, devised by Francis Brown~\cite{algo},
is capable of re-deriving this data. The method is superior
to brute-force LLL or PSLQ, since it requires only a single
rational number to be determined empirically, at each
iteration, the rest of the work being achieved by integer
arithmetic alone. After I had made the data up to $w=31$
available in the MDV datamine, Panzer used Brown's method to
show that the coefficients $C_{w,k}$ must satisfy the sum rules
\begin{equation}
\sum_{k=1}^j{2j-1\choose 2k-1}\left(C_{w,k}-C_{w,k-j+(w-1)/2}\right)
=\frac12(1-2^{2j+1-w})(1-3^{2j+1-w})C_{w,j}
\label{epsr}
\end{equation}
for $j=1,\ldots,(w-3)/2$. After requiring that $C_{w,1}>0$,
I showed that for all odd $w\le601$ these sum rules require
that $C_{w,k}>0$ for $k=1,\ldots,(w-1)/2$, except for the
case $w=273$, remarked on by Panzer, where $C_{273,k}<0$
for $k=52,\ldots,85$, and the case $w=585$, where $C_{585,k}<0$
for $k=41,\ldots,251$. Hans Bethe might have been amused to see
that $273=2\times137-1$, in his satire~\cite{bethe} on Eddington,
is not unique, in the present context.

These remarkable findings for imaginary parts at odd weights
and depth 2, now encapsulated by Conjecture~4, led me to
investigate real parts at even weights and depth 2, with
results now encapsulated by Conjectures~5 and 6, which have
been tested up to weight $w=36$, which is the first weight
for which there are 3 modular forms believed to be relevant to
pushdown. Here my intuition suggested that the rational
numbers would be under much better control, since real parts
in the $\{A,B,D\}$ alphabet include the MZVs of Zagier's
$\{A,B\}$ subalphabet, where bizarre primes with more than
100 digits are not expected at depth 2 and weight $w\le36$.

I took as my study a single word, $A^{w-2}D^2$, at each even
weight $w$, since $\Re Z(A^{w-2}D^2)$ was readily
observed to be an {\em honorary} MZV for even $w<12$, as shown here:
\begin{eqnarray}
\Re Z(D^2)&=&-\frac13\zeta_2\\
\Re Z(A^2D^2)&=&-\frac{23}{216}\zeta_4\\
\Re Z(A^4D^2)&=&\frac{209}{972}\zeta_6-\frac16\zeta_3^2\\
\Re Z(A^6D^2)&=&\frac{799331}{1399680}\zeta_8
-\frac{25}{54}\zeta_5\zeta_3
-\frac{7}{270}\zeta_{5,3}\\
\Re Z(A^8D^2)&=&\frac{31013285}{35271936}\zeta_{10}
-\frac{535}{2016}\zeta_5^2
-\frac{637}{1296}\zeta_7\zeta_3
-\frac{205}{18144}\zeta_{7,3}\end{eqnarray}
where $\zeta_{a,b}\equiv\sum_{m>n>0}1/(m^a n^b)$.

Had one stopped at this point, there would have been little
point in presenting Conjecture~5 as worthy of attention. Surely
such reducibility to MZVs will continue? Indeed it does,
but in a very refined manner, which is consistent with
Conjecture~5 but also engages the modular forms and
alternating sums of Conjecture~6.

A naive empiricist might have expected reducibility of
$\Re Z(A^{10}D^2)$ to the basis ${\rm
MZV}_{12,2}\equiv\{\zeta_{12},\zeta_9\zeta_3,
\zeta_7\zeta_5,\zeta_{9,3}\}$ that is
proven~\cite{mine} to suffice for the reduction of MZVs of
weight 12 and depth 2.
As an author of~\cite{bk}, I did not expect this and was
hence delighted by the absence such of such a reduction.
Rather one needs to adjoin to ${\rm MZV}_{12,2}$ an
alternating sum, such as $Z(A^8C A^2B)=U_{9,3}\equiv\sum_{m>n>0}
(-1)^{m+n}/(m^9n^3)$.

The reason is clear: at $w=12$ there is a depth-4 MZV
that is not reducible to MZVs of lesser depth but is pushed
down to $U_{9,3}$ in the $\{A,B,C\}$ alphabet. It makes no
sense to adjoin a depth-4 MZV to ${\rm MZV}_{12,2}$,
without also adjoining its miscellaneous entourage of
tedious products of primitives, such as
$\zeta_5\zeta_4\zeta_3$ in Equation~(26) of~\cite{alt}. It makes
perfect sense to adjoin solely $U_{9,3}$, to ensure
reducibility of $\Re Z(A^{10}D^2)$.

At $w=14$, the BK conjecture~\cite{bk} asserts that there is no such pushdown,
and indeed a basis for $\Re Z(A^{12}D^2)$ is provided by
a basis ${\rm MZV}_{14,2}$ for MZVs of weight 14 and depth 2:
\begin{eqnarray}6^{10}\Re Z(A^{12}D^2)&=&
\frac{45336887777}{594}\zeta_{14}
-30203052\zeta_{11}\zeta_{3}
-\frac{292990340}{11}\zeta_9\zeta_5\\&&{}
-\frac{400333213}{33}\zeta_7^2
+\frac{19112030}{33}\zeta_{11,3}
-\frac{1938020}{9}\zeta_{9,5}.
\end{eqnarray}
Thereafter one will, by Conjecture~6, always need to adjoin
at least one alternating sum. At $w=24$ two alternating sums
are needed, according to Conjecture~6, as is confirmed by
{\tt lindep}. At $w=36$, three alternating sums are predicted to
be needed and are indeed found to be present.

The story up to weight 36 is contained in the
following delectable list of rational data:\\
{\tt
[12, [256]]\\{}
[16, [19840]]\\{}
[18, [184000]]\\{}
[20, [1630720]]\\{}
[22, [14728000]]\\{}
[24, [165988480, 10183680]]\\{}
[26, [51270856000/43]]\\{}
[28, [13389295360, 808012800]]\\{}
[30, [1573506088000/13, 96652800000/13]]\\{}
[32, [1085492600192, 65740846080]]\\{}
[34, [3003044404360000/307, 182805638400000/307]]\\{}
[36, [95110629053440, 8048874470400, 410097254400]]}\\
where the first entry in each line is the weight $w$
and thereafter I give the unique vector of rational
numbers, $Q_{w,k}$, whose existence was asserted in Conjecture~6.

To each set of modular forms of weight $w\le36$, I have thus
empirically associated a set of eminently printable
rational numbers, with the same cardinality as the modular forms.
It seems to me that a derivation of this data by {\em exact} methods
ought to be within the reach of our wide community.

\section{The lure of Lyndon}

{\em Dare, and the world always yields: or, if it beat you
sometimes, dare again, and it will succumb}, said Barry
Lyndon, the eponymous anti-hero of a novel by William
Makepeace Thackeray (1811-1863).

In 1954, Roger Lyndon (not to be confused with Thackeray's
rogue) defined a subset of words that has remarkable utility
in a wide range of problems, including Lie algebras and
shuffle algebras. Suppose that we impose on an alphabet an
ordering of letters, saying for example that $A$ comes
before $B$, which comes before $C$, etc, or in the case that
we happen to represent letters by positive integers, that 1
comes before 2, which comes before 3, etc. Then we may
define a lexicographic ordering of words in that alphabet,
saying, for example, that $ABACA$ comes before $BCC$, as in
a dictionary. Then a Lyndon word is a word $W$ such that for
every splitting $W=UV$ we have $U$ coming before $V$. Thus
$ABACA$ is not a Lyndon word, because $ABAC$ does not come
before $A$, and $BCC$ is a Lyndon word, because $B$ comes
before $CC$, and $BC$ comes before $C$. The present
significance of this definition is that every shuffle algebra
may be solved by taking the Lyndon words as primitive.

In the case of alternating sums, in the alphabet
$\{A,B,C\}$, we have two closed algebras, the shuffles of iterated
integrals and the stuffles of nested sums, whose stuff may be
ignored, for present purposes. Then Lyndon immediately tells
us how to solve one of these, say the shuffles,
leaving us to tussle with the stuffles, along with, perhaps,
further relations, such as the doubling relations and word
transformation explained in~\cite{mine}. Here, Lyndon cannot
tells us what to do, since we have already used him once.
Yet he may still guide us, if we happen to have a simple
conjecture for the enumeration of primitives that remain
after all our relations have been satisfied and are able to spot
a Lyndon-type prescription with the same enumeration.

That was precisely the situation in which I found myself in
the case of alternating sums, after conjecturing~\cite{alt} the
enumeration $E_{w,d}=T((w-d)/2,d)$, mentioned in Section~2.
This coincides with the number of Lyndon words in an
alphabet of {\em odd} integers, with the weight
corresponding to the sum of the integers and the
depth to the length of the word. Then it was easy
to guess a set of primitives.

In the present case of MDVs, there is a comparable
situation. There is good reason to trust the enumeration
$N_{w,d}=T(w-d,d)$ that follows from Conjecture~2. It is
easily shown that this coincides with the number of Lyndon
words in an alphabet of integers greater than 1, with the
weight corresponding to the sum of the integers and the
depth to the length of the word. Combining this
observation with the generalized parity
conjecture~\cite{sixth}, one may write
down a list of 53 Lyndon {\em symbols} that
act as placeholders for the primitives with weights up to 11.
At depth 1, one may write:
\begin{equation}I_2, R_3, I_4, R_5, I_6, R_7, I_8, R_9, I_{10}, R_{11},\end{equation}
at depth 2:
\begin{equation}
I_{2,3},
R_{2,4},
I_{2,5}, I_{3,4},
R_{2,6}, R_{3,5},
I_{2,7}, I_{3,6}, I_{4,5},
R_{2,8}, R_{3,7}, R_{4,6},
I_{2,9}, I_{3,8}, I_{4,7}, I_{5,6},\end{equation}
at depth 3:
\begin{eqnarray}
R_{2,2,3},I_{2,2,4}, I_{2,3,3},R_{2,2,5}, R_{2,3,4}, R_{2,4,3},I_{2,2,6}, I_{2,3,5}, I_{2,4,4},&&\\
I_{2,5,3}, I_{3,3,4},R_{2,2,7},R_{2,3,6}, R_{2,4,5}, R_{2,5,4},R_{2,6,3},R_{3,3,5},R_{3,4,4},
\end{eqnarray}
at depth 4:
\begin{equation}
I_{2,2,2,3},
R_{2,2,2,4}, R_{2,2,3,3},
I_{2,2,2,5}, I_{2,2,3,4}, I_{2,3,2,4}, I_{2,2,4,3}, I_{2,3,3,3},\end{equation}
and, finally, at depth 5,
\begin{equation}R_{2,2,2,2,3}.\end{equation}

Commitments to precise choices for the depth-1 primitives
and the imaginary depth-2 primitives were made in Section~4,
with sound reasons given there. The remaining 33 choices
will be made in Sections~6 and~8. In the rest of this section,
I shall discuss an alluring temptation that has
been firmly resisted in this paper.

The temptation is to model a basis for MDVs
on the basis that I successfully devised for
alternating sums. In the $\{A, B, C\}$ alphabet,
I discovered that the primitives may be
taken to be words in the subalphabet $\{A, C\}$,
where they are hence words of the form
$A^{n_1-1}C A^{n_2-1}C\ldots A^{n_d-1}C$
at depth $d$ and weight $w=\sum_j n_j$.
If one demands that the $n_j$ are {\em odd} integers
and form a vector whose reverse is a Lyndon word,
then a viable basis results, as I informed Deligne,
Ihara and Zagier, in May 1997~\cite{letterd},
providing what Deligne has described as
{\em une \'evidence num\'erique \'ecrasante}~\cite{PD1}.

So it seemed natural, both to Panzer and to Schnetz, that a
sound basis for MDVs would be achieved by using the
words $A^{n_1-1}D A^{n_2-1}D\ldots A^{n_d-1}D$ with integers
$n_j>1$ whose reversed vector is a Lyndon word in the
alphabet whose letters are the
integers {\em greater than unity}~\cite{PD1}.
Then by taking the real parts when the depth and weight have the
same parity, and the imaginary parts when those parities
differ, one has a perfect fit to the enumerations of
Conjectures 1, 2 and 3. I shall refer to this as the {\em
Deligne basis} for MDVs, since in response to my
study of the $\{A,B,C\}$ alphabet Pierre wrote: ``If
$\lambda=\frac12(1+\sqrt{-3})$ (sixth root of 1), one could
hope for having a similar story\ldots" as I gratefully
recorded in Section 5.4 of~\cite{sixth}.

It was by choosing this Deligne basis that Panzer and
Schnetz obtained the disturbing primes 50909 and 121577 in the
{\em denominators} of their rational coefficients of
reduction.

First, let's see how 50909 infected the Feynman period.
At weight 9 and depth 2,
\begin{equation}X_9=2592I_{1,8}+2538I_{2,7}+2607I_{3,6}+1318I_{4,5}
=\frac{1357169441}{5}P_9\end{equation}
is the sole relation between imaginary parts that is not
obtainable by trivial algebra restricted to depth 2. Note well
that $X_9$ causes no inconvenience whatsoever to the datamine of
Section~8, which is already committed to eliminating
$I_{1,8}$, whose coefficient $2592=2^5 3^4$ is quite
harmless, as a divisor.

Now suppose that one elects to use the Deligne basis.
Consider the imaginary parts $J_{a,b}\equiv\Im Z(A^{b-1}D A^{a-1}D)$,
with $b>a>0$ and odd weight $a+b$.
The Deligne basis regards these as primitive when $a>1$,
but not when $a=1$. It is very easy to transform the
set $\{I_{1,8},I_{2,7},I_{3,6},I_{4,5}\}$ in $X_9$
to the corresponding set in the $J$-language. Of course
that will introduce products of depth-1 primitives,
from which $X_9$ is notably free, but these are not
at issue at present; what we care about here are the integer
coefficients in the reducible combination of $J$'s.
So let us abandon equality, pro tempore, and content ourselves
with using the symbol $\sim$ to indicate that
we are working modulo products and powers of $\pi$.
Then simple shuffles and
stuffles at depth 2 tell us how to
record, in $J$-language, the inescapable fact that
$X_9$ is reducible. Here is the result
\begin{equation}
50909J_{1,8}+25020J_{2,7}+10083J_{3,6}+2538J_{4,5}\sim0
\end{equation}
whence came the infection by the denominator prime 50909
in the work of both Panzer and Schnetz, who had adopted
the Deligne basis, in which one has pledged, in advance,
to eliminate $J_{1,8}$.

Similarly, at weight 11, the reducibility of $X_{11}$
translates, in their $J$-language, to
\begin{eqnarray}
6239210063J_{1,10}+3133054680J_{2,9}+1337436381J_{3,8}&&\\{}
+443069676J_{4,7}+ 87845202J_{5,6}&\sim&0
\end{eqnarray}
and hence they were, albeit unwittingly, committed to
dividing by $6239210063=19\times37\times73\times121577$,
which circumstance explains the large prime 121577 noted in
the abstract. At $w=13$, the elimination of $J_{1,12}$
produces two large denominator primes: 10137187 and
216364363. Moreover the destructive effect of such
denominator primes is cumulative: at weight 11, the
denominator of the coefficient of $\pi^{11}$ of a reduction
of the imaginary part of a depth-4 MDV to the Deligne basis
will, generically, contain the primes 19, 37, 73, 50909, and
121577. For the imaginary parts of depth-6 MDVs of weight
13, those primes will, in general, be joined by 10137187 and
216364363. In consequence, one must be prepared to
encounter huge numerators of the coefficients of
reduction to the Deligne basis. This is disastrous when
attempting to fit a numerical result
using the LLL or PSLQ algorithms, which will require much
greater precision than would have been needed had one forsaken
the Deligne basis for a more practical one.

In retrospect, it is now clear why Panzer needed to increase
his numerical precision from 1000 to 5000 digits before
finding a fit to the Feynman period: the Deligne basis is
extremely unfriendly to empiricists.

\section{Taming the basis}

My Aufbau for the basis of the datamine of Section~8 is unashamedly
empirical.
I was determined to explore {\em all} of the relations
between MDVs up to weight 11, starting at low weight and working
my way up, by weight. Moreover, at each weight, I started at low depth
and worked my way up, by depth.

There is no more to say here about weights
$w<6$, since the die is already cast for the depth-1 primitives
and the imaginary primitives of depth 2. At weights 6 and 7,
no harm is done by setting $R_{2,4}=\Re Z(ADA^3D)$ and
$R_{2,2,3}=\Re Z(ADADA^2D)$. At weight 8, the depth-3
choices $I_{2,2,4}=\Im Z(ADADA^3D)$ and
$I_{2,3,3}=\Im Z(ADA^2DA^2D)$ are likewise unobjectionable.
An attentive reader may have noticed that my powers of $A$,
thus far, were precisely the {\em reverse} of those for the Deligne
basis. Life becomes more interesting at $w=8$ and $d=2$, where
we meet an irreducible MZV. Here I chose
$R_{2,6}=\Re Z(A^5DAD)$ and $R_{3,5}=Z(A^5BAB)$,
the latter being a MZV. I remind the reader that the indices
of primitives are purely formal: they imply no prior commitment
to specific powers of $A$ or to the presence or absence of $B$.

Weight 9 was a little trickier. There was no objection
to $I_{2,2,2,3}=\Im Z(ADADADA^2D)$ but effort
was needed to determine that
\begin{equation}
R_{2,2,5}=\Re Z(A^2DA^4D^2),\quad
R_{2,3,4}=\Re Z(ADA^2DA^3D),\quad
R_{2,4,3}=\Re Z(A^5D^2AD)
\end{equation}
avoid unwanted denominator primes. Here the strategy
was empirical: choose 3 words of weight 9 and depth
3, more or less at random; if {\tt lindep} reveals that a
combination of them is reducible, then go back and choose another
triple, else take this triple as a part of a {\em temporary}
basis to which all such words of this weight and depth
may be reduced; perform those
reductions and throw away all products and powers of $\pi$;
look for unwanted denominator primes; if some are found
remove them, where possible, by studying the determinants of
$3\times3$ matrices relating alternative triples to
those of the temporary basis. This may give a {\em refined}
basis at the current depth. But we are still not done. Study
reductions of MDVs of the same weight, but greater depth, to
the refined basis and determine whether new primes appear.
If they do, go back and try a different refinement.

I shall refer to such a strategy as {\em trial and error}.
Conjecture~4 indicates that this is bound to fail,
eventually. Then I resort to a method of {\em modular rectification}
illustrated below at weights 10 and 11.

At weight 10, trial and error took a good deal of effort,
which was rewarded by the successful choices
\begin{eqnarray}
I_{2,2,6}=\Im Z(A^7D^3),\quad
I_{2,3,5}=\Im Z(DBA^7D),\quad
I_{2,4,4}=\Im Z(ADA^2DA^4B),&&\\{}
I_{2,5,3}=\Im Z(A^2DA^2DA^3D),\quad
I_{3,3,4}=\Im Z(A^2DA^2DA^3B)
\end{eqnarray}
at depth 3, where I left the $\{A,D\}$
subalphabet to achieve refinement.
At depth 4, the simplistic choices
$R_{2,2,2,4}=\Re Z(ADADADA^3D)$ and
$R_{2,2,3,3}=\Re Z(ADADA^2DA^2D)$
proved harmless. At depth 2, the choices
\begin{equation}
R_{2,8}=\Re Z(ADA^7D),\quad
R_{3,7}=\Re Z(A^5DA^3D),\quad
R_{4,6}=Z(A^7BAB)
\end{equation}
are a good refinement, but then reductions from greater
depths produce the denominator prime 43, which I did not
succeed in removing by a new refinement. However that is not a
problem. I computed the reductions of {\em all} of the
26,244 finite MDVs of weight 10, multiplied by 43 and
then took residues modulo 43. This showed that a {\em single}
transformation
\begin{equation}
R_{2,8}=43T_{2,8} +2R_{3,7} - 9R_{4,6} - 5R_3R_7 - 6I_4I_6
+ (\pi/3)^{10}/11!
\end{equation}
removes the unwanted prime 43 at weight 10. So the
MDV datamine uses $T_{2,8}$ instead of $R_{2,8}$.
Note that, in contrast with trial and error, modular rectification
requires one to retain products and powers of $\pi$ in the
reductions to the unrefined basis. Moreover it is not known
in advance whether a single transformation will be
effective in removing all the poles at a given prime.
When several primes must be removed, a transformation may
(but need not necessarily) be required for each.

At weight 11, the choice $I_{2,2,3,4}=\Im Z(A^5DA^2D^3)$ was useful.
Then trial and error yielded
\begin{eqnarray}
I_{2,2,2,5}=\Im Z(A^5D^2ADAD),\quad
I_{2,3,2,4}=\Im Z(A^3DADADA^2D),\quad&&\\
I_{2,2,4,3}=\Im Z(A^3DADA^3D^2),\quad
I_{2,3,3,3}=\Im Z(A^2DA^2DA^2DAD)
\end{eqnarray}
as a successful refinement for depth 4.
Thus we are almost done, as far
as the Feynman period is concerned, since that does not involve
the real parts at weight 11, which are postponed to
Section~8.

It remains to address the ineluctable difficulty posed by
the success of Conjecture 4, whose faithful witness, $X_{11}$, at
weight 11 is
\begin{eqnarray}
79816752I_{1,10}
+84001536I_{2,9}
+87845202I_{3,8}&&\\{}
+80697891I_{4,7}
+40070327I_{5,6}&=&841838813449645P_{11}.
\end{eqnarray}
Here the problem is less severe than in the Deligne
basis, since my commitment to eliminate $I_{1,10}$
introduces only a 4-digit prime, namely
$2281=79816752/(2^4 3^7)$. As in the case of weight 10,
the prime 43 also creeps down from depths $d>2$ .

So now the method is clear: compute the reductions
of the imaginary parts of the 78,732 MDVs
of weight 11 and determine from the residues of their
poles, modulo 43 and modulo 2281, whether two transformations
suffice to remove both primes from the datamine. Happily,
this modular rectification
\begin{eqnarray}
I_{2,9}&=&91(11T_{2,9}) - 898T_{3,8}
+ 11I_{4,7} - 292P_{11}\\
I_{3,8}&=&24(11T_{2,9}) + 841T_{3,8}
- 190I_{4,7} - 255P_{11}
\end{eqnarray}
suffices, with a determinant $91\times841+24\times898=43\times2281$
that neatly solves the problem of 43 and 2281. As a bonus,
inclusion of 11 in the multiples of $T_{2,9}$ renders
reductions of the imaginary parts of weight-11 MDVs in the
$\{A,D\}$ alphabet free of the denominator prime 11, up to
depth 4. So now we are ready for the Feynman period.

\section{The Feynman period $P_{7,11}$}

All imaginary parts of MDVs with weight 11 and depth $d\le4$
are reducible to a sub-basis of dimension of 57; the
additional 15 terms in the full basis appear only in MDVs
with $d>4$. If one gives 1050 digits of $\sqrt{3}P_{7,11}$
to {\tt lindep} and asks for a reduction to the 57-dimensional
sub-basis of the datamine, a valid answer is returned,
agreeing with Panzer's 5000-digit result. The MDV datamine
then enables one to investigate whether the terms of depth 4
have any distinctive pattern. They do indeed.

In the datamine basis,
the depth-4 contribution is a linear combination of
$\Im Z(A^5DA^2D^3)$
and 6 products terms, {\em all} of which contain $\zeta_3$.
Then a beautiful thing happens if one uses the
datamine to transform to $\Im Z(A^7D^4)$. Now one
simply obtains the depth-4 term as an integer
multiple of $\Im Z(W_{7,4})$ with a word-combination
\begin{equation}
W_{7,4}\equiv A^7D^4+\zeta_3A^5D^3
+\frac12\zeta_3^2A^3D^2+\frac16\zeta_3^3AD\label{w74}
\end{equation}
having the formal appearance of a Taylor expansion in $\zeta_3$
in which taking a derivative corresponds to removing the sub-word
$AAD$. For $m\ge2(n-1)$, we may adopt this
general definition:
\begin{equation}
W_{m,n}\equiv\sum_{k=0}^{n-1}
\frac{\zeta_3^k}{k!}A^{m-2k}D^{n-k}.\label{wmn}
\end{equation}
Then, wonderful to relate,
the depth-3 terms involve only $\pi^2\Im Z(W_{7,2})$
and MZVs multiplied by $\pi$, as shown here:
\begin{eqnarray}
\sqrt{3}P_{7,11}&=&
\label{p1}
-10080\Im Z(W_{7,4}+W_{7,2}P_2)+50400\zeta_3\zeta_5P_3\\&+&
\label{p2}
\left(35280\Re Z(W_{8,2})+\frac{46130}{9}\zeta_3\zeta_7
+17640\zeta_5^2\right)P_1\\&-&
\label{p3}
13277952T_{2,9}-7799049T_{3,8}
+\frac{6765337}{2}I_{4,7}-\frac{583765}{6}I_{5,6}\\&-&
\label{p4}
\frac{121905}{4}\zeta_3I_8-93555\zeta_5I_6
-102060\zeta_7I_4-141120\zeta_9I_2\\&+&
\label{p5}
\frac{42452687872649}{6}P_{11}
\end{eqnarray}
where, as usual, $P_n\equiv(\pi/3)^n/n!$.

{\bf Remarks:} On the first line we see that
$W_{7,2}\equiv A^7D^2+\zeta_3A^5D$ combines with $W_{7,4}$
in the simplest manner imaginable. On the second line,
$W_{8,2}\equiv A^8D^2+\zeta_3A^6D$ gives MZVs,
since $\Re Z(A^8D^2)$ is an honorary MZV of
depth 2, in accord with Conjecture 6, and $\Re Z(A^6D)$ is
a rational multiple of $\zeta_7$.
Thanks to the use of the datamine transforms $T_{2,9}$ and
$T_{3,8}$ on the third line, no prime greater than 3 appears
in any denominator. On the final line we see that the
numerator of the coefficient of $\pi^{11}$ is
$42452687872649=31\times1369441544279$, with 14 digits.
Panzer had a 30-digit numerator and Schnetz obtained
a 50-digit numerator in the course of investigating
a very interesting coaction conjecture for Feynman
periods~\cite{letter}.

\section{Structure of the MDV datamine}

For the real parts at weight 11,
I began with the Deligne basis:
$R_{2,2,7} =\Re Z(A^6DADAD)$,
$R_{2,3,6} =\Re Z(A^5DA^2DAD)$,
$R_{2,4,5} =\Re Z(A^4DA^3DAD)$,
$R_{2,5,4} =\Re Z(A^3DA^4DAD)$,
$R_{2,6,3} =\Re Z(A^2DA^5DAD)$,
$R_{3,3,5} =\Re Z(A^4DA^2DA^2D)$,
$R_{3,4,4} =\Re Z(A^3DA^3DA^2D)$
and $R_{2,2,2,2,3} =\Re Z(A^2DADADADAD)$.
This is, of course, very inefficient. I found that
it introduces the denominator primes
47, 71 and 19766363. By computing the residues
of poles at these primes, I found a pair
of transformations, from $\{R_{2,6,3},R_{3,4,4}\}$
to rectified datamine primitives $\{T_{2,6,3},T_{3,4,4}\}$ that
remove those primes from the denominators.

The datamine
{\tt http://physics.open.ac.uk/$\widetilde{\phantom.}$dbroadhu/cert/MDV.tar.gz}
contains 24 files, which have been tarred and zipped
for downloading as a single 84-megabyte file. After opening this,
a user should consult {\tt readme.txt}, which lists the 24 files.

{\tt MDVdef.txt} defines 53 primitives, to which
5 transformations, recorded in {\tt MDVtra.txt}, are applied,
to determine the symbols used in the basis file {\tt MDVbas.txt},
which also uses the symbol {\tt p3} for $\pi/3=\Im Z(D)$.
Rational data for the MDVs are contained in
10 files, organized by weight, and in the case of weight 11
also by depth. Each of the 118,097 lines in these files
gives a word and a pair of vectors of rational coefficients.
Thus, for example, the entries
\begin{verbatim}
[[A, B, D, B], [[-9/8, 201/20], [-27/8, 0, 33/4]]]
[[I2^2, 1/24*p3^4], [I4, R3*p3, 1/2*I2*p3^2]]
\end{verbatim}
in {\tt MDVw7.txt} and {\tt MDVbas.txt} record the reduction
\begin{equation}
Z(ABDB) = -\frac98I_2^2+\frac{201}{20}\frac{(\pi/3)^4}{4!}
-\frac{27i}{8}I_4+\frac{33i}{4}I_2\frac{(\pi/3)^2}{2!}
\end{equation}
and one may use evaluations of primitives
in {\tt MDVpri.txt} to obtain up to 20000 digits of
this MDV. This has been conveniently automated,
for unix users of {\tt Pari-GP}, by a
utility file {\tt MDVgrep.gp}. Here is a very simple
example of its use
\begin{verbatim}
? \r MDVgrep.gp
? default(realprecision, 25);
? print(grep(ABDB))
-0.01147228319232732872517521 + 0.3132313899574234502633605*I
\end{verbatim}
which works by issuing an {\tt extern} command to find the
relevant line of rational data. Hence it is not necessary to
load the full datamine into memory; only a single megabyte
of numerical data for the primitives is needed. Yet that
megabyte provides speedy access to about 5 gigabytes of numerical values
for all MDVs up to weight 11.

{\tt MDVconj4.txt} has data on Conjecture 4, up to $w=31$,
and {\tt MDVconj6.txt} has data on Conjecture 6, up to $w=36$.
{\tt MDVprime.txt} provides a list of primes
sufficient to factorize the large integers in
{\tt MDVconj4.txt}, as is done by {\tt MDVfact.gp},
with results in {\tt MDVfact.out}.

{\tt MDVtest.gp} tests the definition file and
{\tt MDVfeyn.gp} tests the formula for the Feynman period $P_{7,11}$
against the 20000-digit value stored in {\tt P7\_11.txt}.
Unix users are advised to run these two tests, to check
that the datamine is being correctly accessed on their operating
system. Windows users
are left high and dry by the inability of their system to
respond to the {\tt extern} command that greps the relevant
line for a given word.

\section{Comments and conclusions}

Adapting a remark by Pliny the Elder,
I observe {\em ex QFT semper aliquid novi}.
It was a 3-loop radiative correction, to the relation
between the Weinberg angle and the masses of the W and Z bosons
of the standard model, that led me to intensive investigation
of weight-4 polylogs of the sixth root of unity in~\cite{sixth},
which resulted in Conjectures 1, 2 and 3. Now weight-11 polylogs from the
restricted alphabet $\{A,B,D\}$ of MDVs have
emerged as contributors to the renormalization of QFT at 7 loops.
The standard model includes a $\phi^4$ term, as the
quartic self-coupling of the Higgs boson, for which at 7 loops
we now know all the counterterms from subdivergence-free diagrams
and hence may be confident that the period $P_{7,11}$ of Figure~2,
given in tolerably compact form by expressions~(\ref{p1}-\ref{p5}),
contributes to the beta-function, in a scheme-independent manner,
since the other subdivergence-free 7-loop diagrams evaluate
to MZVs~\cite{BS}.

In the course of obtaining a formula
for $P_{7,11}$ that is free of any denominator prime greater than 3,
I was led to the rather specific Conjectures 4, 5 and 6 and to an Aufbau
that provides a datamine of 13,369,520 rational coefficients,
free of denominator primes greater than 11, from which one may now
speedily obtain 20000 good digits of the real and imaginary parts of
all of the 118,097 MDVs with weights up to 11, using a
handy utility {\tt MDVgrep.gp} provided with the
MDV datamine.

I have formulated all but one of Conjectures 1 to 6 in a form strong enough
to make proof within my lifetime improbable.
The exception is Conjecture~5, which
may be accessible to a method that Claire Glanois has developed to
determine classes of alternating sums that yield honorary MZVs.

In conclusion, I offer the following observations.
\begin{enumerate}
\item MDVs are radically different from alternating sums
in the $\{A,B,C\}$ alphabet,
since the $\{A,B,D\}$ alphabet of MDVs is not closed under stuffles.
\item Conjecture~4 asserts
the existence of reducible combinations of depth-2 imaginary parts
that cannot be reduced by algebra restricted to depths $d\le2$.
\item To compensate for the paucity of useful stuffles, MDVs
are endowed with the powerful relation of Lemma~1, whereby
the complex conjugate of a MDV is, up to a sign, a MDV of
different depth, in general.
\item This latter feature permits a divide-and-conquer splitting
of the Fibonacci numbers, by the real and imaginary generating
functions~(\ref{DR},\ref{DI}),
thereby extending the reach of integer-relation searches by the
LLL and PSLQ methods.
\item This advantage was negated in the work of Panzer and
Schnetz by their adoption of a Deligne basis that generates
gratuitously large primes in denominators.
\item Such denominator primes are avoided in the MDV
datamine of Section~8, whose construction
at weights 10 and 11 took a combination of trial and error with a method
of modular rectification, exemplified in Section~6.
\item My simplification in Section~7 of Panzer's result for the
counterterm for the Feynman diagram in Figure~2
depended crucially on the new datamine, which revealed
a notable Taylor-like expansion~(\ref{w74}) at depth 4. This was
generalized in~(\ref{wmn}) to clean up, in similar fashion,
the depth-3 contributions in expressions~(\ref{p1}-\ref{p5}).
\item In the spirit of Grothendieck, Deligne~\cite{PD,PD1} and
Brown~\cite{FB,FB1},
it is motivically possible to prove a weakened
version of Conjecture 2,
obtaining upper bounds for the numbers of primitive MDVs, graded
by both weight and depth. Here, one would dearly like
to have an explanation of why my simple-minded route to
the generating function~(\ref{myc}) does {\em not} work
for MZVs of the subalphabet $\{A,B\}$,
where the BK conjecture~(\ref{bkc})
insists on an extra term, which also enumerates
modular forms at even weights $w=12$ and $w>14$.
\item Conjecture~6 asserts that a single MDV assigns
a unique set of rational numbers to a set of modular
forms with the same cardinality. This seems to me to be worthy
of further investigation.
\end{enumerate}

{\bf Acknowledgments}:
I owe much to my co-authors,
Johannes Bl\"umlein,
Jonathan Borwein,
David Bradley,
John Gracey,
Dirk Kreimer,
Petr Lisonek,
Oliver Schnetz,
and Jos Vermaseren, for past help with MZVs and
alternating sums.
Encouragement from discussions in Berlin with Freeman Dyson,
in May 2014, on number theory, QFT and the poetry of
George Herbert (1593-1633),
gave me a fine stimulus to the efforts reported here.
I thank Stephen Broadhurst for guiding me from the Fibonacci
numbers to Pascal's triangle in~(\ref{stephen}),
in the mid 1990s, when he was yet a schoolboy.
Patient advice, from
Spencer Bloch,
Francis Brown,
Pierre Deligne,
Stephen Lichtenbaum
and
Don Zagier,
has helped me to appreciate
both the power and also the limitation
of pure mathematics, unaided by empiricism.
My chief debts, in this paper, are to my
supervisor, Gabriel Barton, who
instilled in me, 45 years ago,
an abiding respect for the role
of analysis in quantum field theory,
and, much more recently, to Erik Panzer, without whose
youthful vigour and firm command of
detail the present investigation
would not have been contemplated.

\raggedright


\begin{thebibliography}{99}

\bibitem{bethe}
G.~Beck, H.~Bethe, W.~Riezler,
{\em Bemerkung zur Quantentheorie der Nullpunktstemperatur},
Naturwissenschaften~{\bf 19} (1931) 39;
translation in
{\em Selected Works of Hans A.~Bethe},
World Scientific Series in 20th Century Physics,
Volume 18 (1997) p. 186,
{\tt http://books.google.com/books?id=5baAG1WqgYQC\&q=273+perfect} .

\bibitem{mine}
J.~Bl\"umlein, D.J.~Broadhurst, J.A.M.~Vermaseren,
{\em The multiple zeta value data mine},
Comput.~Phys.~Commun.~{\bf 181} (2010) 582-625,
[arXiv:0907.2557].

\bibitem{bbb}
J.M.~Borwein, D.M.~Bradley, D.J.~Broadhurst,
{\em Evaluations of $k$-fold Euler/Zagier sums:
a compendium of results for arbitrary $k$},
Electron.~J.~Combin.~{\bf 4} (1997) R5,
[arXiv:hep-th/9611004].

\bibitem{bbbl}
J.M.~Borwein, D.M.~Bradley, D.J.~Broadhurst, P.~Lisonek,
{\em Combinatorial aspects of multiple zeta values},
Electron.~J.~Combin.~{\bf 5} (1998) R38,
[arXiv:math/9812020].

\bibitem{ams}
J.M.~Borwein, D.M.~Bradley, D.J.~Broadhurst, P.~Lisonek,
{\em Special values of multiple polylogarithms},
Trans.~Amer.~Math.~Soc.~{\bf 353} (2001) 907-941,
[arXiv:math/9910045].

\bibitem{alt}
D.J.~Broadhurst,
{\em On the enumeration of irreducible $k$-fold Euler sums
and their roles in knot theory and field theory},
[arXiv:hep-th/9604128].

\bibitem{letterd}
David Broadhurst,
{\em Tests of the enumeration $E_{n,k}$},
letter to Deligne, Ihara, Zagier, et al,
May 1997.

\bibitem{sixth}
D.J.~Broadhurst,
{\em Massive 3-loop Feynman diagrams reducible to {\tt SC$^{*}$}
primitives of algebras of the sixth root of unity},
Eur.~Phys.~J.~{\bf C8} (1999) 311-333,
[arXiv:hep-th/9803091].

\bibitem{bgk}
D.J.~Broadhurst, J.A.~Gracey, D.~Kreimer,
{\em Beyond the triangle and uniqueness relations:
non-zeta counterterms at large $N$ from positive knots},
Z.~Phys.~{\bf C75} (1997) 559-574,
[arXiv:hep-th/9607174].

\bibitem{pisa}
D.J.~Broadhurst, D.~Kreimer,
{\em Knots and numbers in $\phi^4$ theory to 7 loops and beyond},
Int.~J.~Mod.~Phys.~{\bf C6} (1995) 519-524,
[arXiv:hep-ph/9504352].

\bibitem{bk}
D.J.~Broadhurst, D.~Kreimer,
{\em Association of multiple zeta values with positive
knots via Feynman diagrams up to 9 loops},
Phys.~Lett.~{\bf B393} (1997) 403-412,
[arXiv:hep-th/9609128].

\bibitem{BS}
David Broadhurst, Oliver Schnetz,
{\em Algebraic geometry informs perturbative quantum field theory},
Proc.~Sci.~{\bf 211} (2014) 078,
[arXiv:1409.5570].

\bibitem{route}
F.C.S.~Brown,
{\em The massless higher-loop two-point function},
Commun.~Math.~Phys.~{\bf 287} (2009) 925-958,
[arXiv:0804.1660].

\bibitem{route1}
F.C.S.~Brown,
{\em On the periods of some Feynman integrals},
[arXiv:0910.0114].

\bibitem{algo}
Francis Brown,
{\em On the decomposition of motivic multiple zeta values},
[arXiv:1102.1310].

\bibitem{FB}
Francis Brown,
{\em Mixed Tate motives over ${\bf Z}$},
Ann.~Math.~{\bf 175} (2012) 949-976,
[arXiv:1102.1312].

\bibitem{FB1}
Francis Brown,
{\em Depth-graded motivic multiple zeta values},
[arXiv:1301.3053].

\bibitem{letterp}
Pierre Deligne,
{\em About your ``Conjectured enumeration of irreducible
Multiple Zeta values$\ldots$''},
letter to the author,
May 1997.

\bibitem{PD}
Pierre Deligne,
{\em Multiz\^etas, d'apr\`es Francis Brown},
S\'eminaire Bourbaki, Jan.~2012, Expos\'e 1048;
Ast\'erisque~{\bf 352} (2013) 161-185,
{\tt http://www.math.ias.edu/files/deligne/012312MultiZetas.pdf} .

\bibitem{PD1}
Pierre Deligne,
{\em Le groupe fondamental de ${\bf G}_m-\mu_N$ pour
$N = 2,~3,~4,~6$ ou $8$},
{\tt http://www.math.ias.edu/files/deligne/121108Fondamental.pdf} .

\bibitem{pam}
M.~Kontsevich, D.~Zagier,
{\em Periods}, in Mathematics Unlimited, 2001 and Beyond
(B.~Engquist and W.~Schmid, eds.), Springer, Berlin
(2001) 771-808,
{\tt http://people.mpim-bonn.mpg.de/zagier/files/periods/fulltext.pdf} .

\bibitem{thesis}
Erik Panzer,
{\em Feynman integrals via hyperlogarithms},
Proc.~Sci.~{\bf 211} (2014) 049,
[arXiv:1407.0074] and thesis recently submitted.

\bibitem{pari}
PARI~Group,
{\tt PARI/GP version 2.5.0}, Bordeaux, 2014,
{\tt http://pari.math.u-bordeaux.fr/} .

\bibitem{census}
Oliver Schnetz,
{\em Quantum periods: a census of $\phi^4$-transcendentals},
Commun.~Number~Theory~Phys.~{\bf 4} (2010) 1-48,
[arXiv:0801.2856].

\bibitem{finite-fields}
Oliver Schnetz,
{\em Quantum field theory over ${\bf F}_q$},
Electron.~J.~Combin.~{\bf 18} (2011) 102,
[arXiv:0909.0905].

\bibitem{letter}
Oliver Schnetz,
{\em $P_{7,11}$ identified},
letter to Bloch, Broadhurst, Brown, Kreimer, Panzer, et al,
May 2014.

\bibitem{oeis}
Neil Sloane,
{\em On-Line Encyclopedia of Integer Sequences},
{\tt http://oeis.org} .

\bibitem{ecm}
Paul Zimmermann, Bruce Dodson,
{\em 20 years of ECM},
{\tt http://www.loria.fr/$\widetilde{\phantom.}$zimmerma/papers/40760525.pdf} .

\end{thebibliography}
\end{document}